\documentclass{svjour3}                     
\smartqed
\usepackage{graphicx}
\usepackage{amsmath}
\usepackage{amssymb}
\journalname{General Relativity and Gravitation}

\begin{document}

\title{Probing Yukawian gravitational potential by Numerical Simulations. I. Changing $N$-Body Codes}

\author{C. S. S. Brandao  \and  J. C. N. de Araujo}

\institute{C. S. S. Brandao \and J. C. N. de Araujo \at
    Divis\~ao de Astrof\'isica \\
    Instituto Nacional de Pesquisas Espaciais  \\
    Avenida dos Astronautas 1758 \\
    S\~ao Jos\'e dos Campos - 12227-010 SP - Brazil \\
              Tel.: +55-12-39457200\\
              Fax:  +55-12-39456811\\
              \email{claudio@das.inpe.br, jcarlos@das.inpe.br}
}

\date{Received: date / Accepted: date}

\maketitle

\begin{abstract}
In the weak field limit general relativity reduces, as is well known, to the Newtonian gravitation. Alternative theories of gravity, however, do not necessarily reduce to Newtonian gravitation; some of them, for example, reduce to Yukawa-like potentials instead of the Newtonian potential. Since the Newtonian gravitation is largely used to model with success the structures of the universe, such as for example galaxies and clusters of galaxies, a way to probe and constrain alternative theories, in the weak field limit, is to apply them to model the structures of the universe. In the present study we consider how to probe Yukawa-like potentials using $N$-body numerical simulations.

\end{abstract}

\section{Introduction}
\label{intro}
Einstein's General Relativity (GR) is one of the most beautiful theories ever imagined by the human mind. Although GR is a successful theory of gravitation, it is unable to explain, for example, the accelerating expansion of the universe, unless a cosmological constant or a dark energy fluid is considered. Nowadays other theories of gravitation intend to give an alternative interpretation to this accelerating expansion (see, e.g., \cite{piazzamarinoni}). Most of these alternative theories, although having different approaches (some scalar-tensor theories of gravitation, nonsymmetric gravitational theory, etc), reduce, in the weak-field limit, to a Yukawa-like gravitational potential (hereafter YGP), i.e.,
\begin{equation}
\phi(r)=-\frac{GM}{r}e^{-r/{\lambda}}.
\label{yukpot}
\end{equation}

The above equation gives the potential of a point mass m at a distance r; G is the universal gravitational constant and $\lambda$ is a constant. When $\lambda \rightarrow \infty$, this potential tends to the Newtonian one. Note that the parameter $\lambda$ is the Compton wavelength of the exchange particle, which in the present case is a graviton. The graviton mass is related to  $\lambda$ through the well known equation $m_g=h / \lambda c$, where $h$ is the Planck constant and c is the speed of light.

The YGP has been investigated in the literature, in particular, in galactic astronomy and cosmology. For example, Signore \cite{signore} studies this potential under the cosmological context, on the other hand, de Araujo \& Miranda \cite{araujoemiranda2007} study how variations of the $\lambda$ parameter can disturb galactic disks. This last study generalizes the (Newtonian) Toomre's \cite{toomre} expressions for rotation curves in order to account for a YGP. Rodriguez-Meza et al. \cite{rodriguezmeza} consider the YGP as a correction on the Newtonian potential.

Recently, some numerical approaches have been developed to study alternative theories. Cervantes-Cota et al. \cite{cervantes2007a,cervantes2007b} have developed numerical studies of a scalar-tensor theory in the weak field limit, using the $N$-body method. In these studies they consider that the gravitational potential in the weak field limit is  given by

$$\Phi_{STT} = \phi_N(r) +  \alpha \phi(r)/ (1+ \alpha)\, , $$

\par\noindent where $\Phi_{STT}$ is the potential from the scalar-tensor theory, $\phi_N(r)$ is the Newtonian potential, $\phi(r)$ is given by Eq.(1), and $\alpha \equiv 1/(3 + 2 \omega)$, with $\omega$ being the Brans-Dicke parameter \cite{brans}. They changed a  particular $N$-body code replacing the Newtonian potential by the potential $\Phi_{STT}$ and analyzed its influence on isolated galaxies, two colliding spiral galaxies and issues concerning the formation of bars. Simulations of cosmological structure formation in $\rm{\Lambda}$CDM scenarios are also studied.

In the present paper, we use similar techniques to that used by Cervantes et al. \cite{cervantes2007a,cervantes2007b}, although with a gravitational potential derived from other theories. We modify a popular $N$-body code, \textbf{Gadget-2} \cite{springel2005}, \textsl{replacing} the Newtonian potential by a pure YGP as given by Equation \ref{yukpot}, which is derived from some alternative theories of gravitation (see, e.g., Visser's theory \cite{visser}).

It is worth stressing that although we have chosen a particular gravitational potential, the approach considered here
can be applied to any alternative gravitational potential.

This paper is organized as follows: in section 2, we show the changes made in the \textbf{Gadget-2} code, in section 3, we test the YGP \textbf{Gadget-2} code using a pair of particles and $N$-body systems. Finally, in section 4, we present the conclusions and briefly discuss how this modified code can be used to test alternative theories of gravitation via galactic dynamics.

\section{Changing the \textbf{Gadget-2} Code}
\label{thecode}

\textbf{Gadget-2} code is a new massively parallel TreeSPH Code, designed to simulate the evolution of $N$-body collisionless systems, like, e.g., cosmological fluids, interacting galaxies, etc. It was developed to work with the concept of particles: matter distribution is represented by a system of particles, where each particle has position, velocity, label, gravitational potential, density (if it represents a gas), and so on.

\textbf{Gadget-2} uses an advanced system of processor's communication, based on MPI interface. MPI is a computer implementation that allows many computers to work in parallel. Thus, one can simulate from thousand of particles using desktops machines to billions of particles using supercomputers. This code computes quickly many gravitational forces using a hierarchical tree algorithm, like the Barnes' Tree Code \cite{barnesehut}, grouping distant particles into larger cells and computing their total potential by multipole expansions. This method is very efficient and the computational effort is $O(N \log N)$ interactions, faster than the $N^{2}$ interactions from direct sum method.  \textbf{Gadget-2} also computes gravitational potentials using particle-mesh (PM) method, when cosmological simulations are considered. In this case, the so called ``long range'' potentials are calculated by the PM method, which computes potentials in the Fourier space using the mesh method. On the other hand, the ``short range'' potentials are calculated in real space using the tree algorithm.

It is worth stressing another very important characteristic of the \textbf{Gadget-2} code: its capability to integrate the equations of motion of particles with a leapfrog integrator based on the the  \textit{KDK scheme} (KDK means kick-drift-kick, see \cite{springel2005}), which conserves the symplectic nature of collisionless systems, in other words, it preserves the Hamiltonian structure of these simulated systems. This approach is more efficient than other ones if applied with the adaptive timestep method, as shown by Springel \cite{springel2005}. In this way, \textbf{Gadget-2} is in general more efficient than codes based on the Runge-Kutta integrators.

It is worth mentioning that \textbf{Gadget-2} offers a set of options: pure newtonian simulations, cosmological comoving periodic simulation with the alternative use of the TreePM formalism, etc. However, the ``Cosmological simulation options'' can be activated or not, whether the code is compiled with these respective options. To simulate only gravitational interactions in local systems, such as galaxies or globular clusters, we compile this code without the advanced cosmological options listed in the Makefile, ``turning on" the gravitation, and ``turning off" all the advanced options, such as PM formalism, expansion of the universe, etc.

In the present paper, we intend to study only gravitational interactions in local systems, without gas and cosmological expansion. To this end, we compile \textbf{Gadget-2} without the advanced options, and so we are able to study a cloud of particles self-interacting gravitationally.

We fully studied the \textbf{Gadget-2} code in order to see where it is necessary to modify it in order to
take into account the corresponding YGP expressions. We concluded that it is only necessary to change the code
where instructions related to the Newtonian potential and accelerations appear. In this case, the acceleration reads
\begin{equation}
\vec{a}= - \frac{G\, M}{\lambda\, r^3}\, e^{- r/\lambda}\, (r+\lambda) \,\vec{r}.
\label{yukacel}
\end{equation}

\par\noindent It is important to note that we have left $\lambda$ as a free parameter.

\section{Testing the Code}
\label{testing}

The reliability of a $N$-body simulation to model a given system depends on how one can control the numerical errors naturally intrinsic in this approach.

For example, the particles have their accelerations computed by a tree recursive code and, consequently, their positions and velocities are updated from the integrators. As a result the total energy of the simulated system, can be well conserved or not, whether the treecode calculations and the integration processes are well accurate or not.
For these reasons, errors  must be investigated and minimized to an acceptable level.

In the $N$-body numerical formalism, one uses the fact that gravitation obeys the \textit{superposition principle}. Therefore, one should firstly test the two-body problem, testing the code's integrator precision to the new gravitational potential one is considering. Subsequent tests should consider $N$-body systems, verifying not only
the treecode precision, but also the influence of some numerical parameters on the timestep values.

\subsection{Simulating the motion of two particles}
\label{binary}

In this first test, we are able to probe how our changes in the \textbf{Gadget-2} code modify its integrator's precision. A similar test was performed by \cite{springel2005} using newtonian physics. This test is very important, because we are interested in verifying the code's precision to drift positions and velocities in each timestep for the YGP.

On the other hand, one could argue that $N$-body systems and a two-body system are very different numerically, and consequently that this test would be useless. However, the $N$-body formalism is based on \textsl{the superposition principle}, in other words, the total potential is the sum of the potential from all individual particles. Then, we firstly test the two-body problem in order to verify if the code can work appropriately. Subsequently, we performe tests with a large amount of particles.

In fact, in the case where treecodes are used,  \textsl{the superposition principle} is used implicitly in the monopole, quadrupole and octopole expansion terms. But,  \textbf{Gadget-2} uses  only the monopolar expansions, which is sufficient and very accurate in the newtonian case, as shown by \cite{springel2005}. In the case where the YGP is used, we  have performed some tests and, as we will show in the next sections, its monopolar term is also enough to yield satisfactory results.

In this first test, we  simulate the motion of two particles: a massive particle and a test particle (whose mass can be neglected). The massive particle has mass M=100.0 (in code units, i.e., M=1.0 $\equiv 1.0 \times 10^{10} M_{\odot}$), and is located at $(0,0,0)$ kpc, the origin of a cartesian coordinate system, and has velocity $\vec{v}=(0,0,0)$  $\rm{km.s^{-1}}$. It is important to note that this system of units is very common in galactic simulations with the \textbf{Gadget-2} code. In this system, the time unit is given in $9.8 \times 10^{8} \sim 10^9$ years.

The test particle has mass m = 0.1, and is located initially at $\vec{r}=(100,0,0)$ kpc with $\vec{v_p}=(0,104,0)$  $\rm{km.s^{-1}}$. We have chosen $\lambda=100$ kpc, for our computational purposes. Note that the initial separation between the particles is equal to $\lambda$. In this simulation, the timestep $\Delta t$ values  are $0.00146484 \le \Delta t \le 0.00585938$. It is worth noting that in \textbf{Gadget-2}, the timestep criterion is

\begin{equation}
\Delta t = min \left[ \Delta t_{max}, \left( \frac{2 \eta \epsilon }{|\vec{a}|} \right) \right],
\label{timestep}
\end{equation}

\par\noindent where $\eta$ is an accuracy parameter; $\Delta t_{max}$, the the maximum timestep allowed, which is usually chosen to be a small fraction of the dynamical time of the system under study; $\epsilon$, the gravitational softening; and  $|\vec{a}|$, the modulus of the particle's acceleration, estimated from the last timestep.
In the subsection 3.1.1, we consider the role of the timestep in our simulations.

\begin{figure}
\begin{center}
\includegraphics[width=10cm]{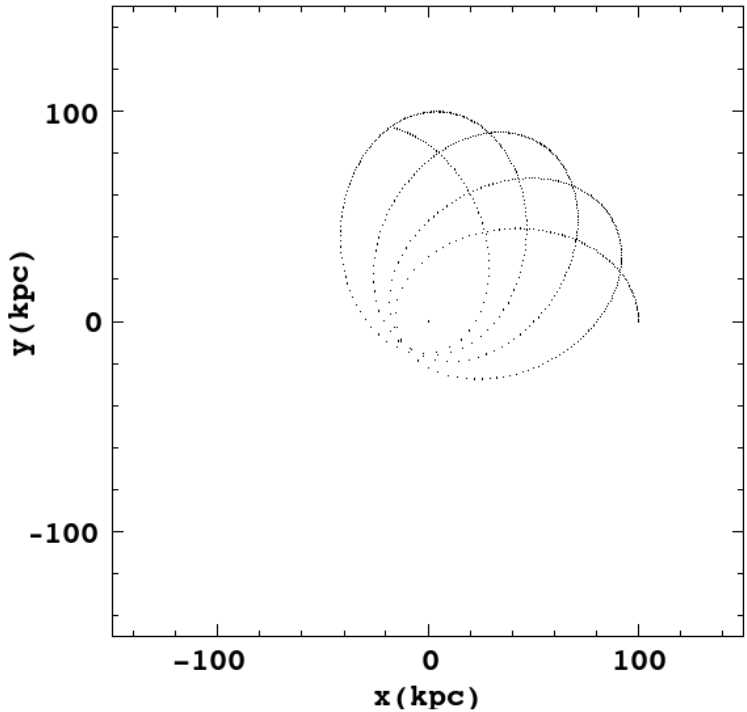}
\end{center}
\caption{Trajectories of the particles on the $xy-$plane. The central particle is the massive one.}
\label{figyuk}
\end{figure}

We have made the system orbiting for 6 Gyears. So, we are able to study the trajectory of the test particle in detail. We display the test particle's orbit in Fig. \ref{figyuk}. This picture is made by displaying the particle's position in the $xy-$plane obtained from 600 snapshots, including the initial one. As can be seen, the test particle orbits in the $xy-$plane and precesses. This orbit characteristic is very important to test the reliability of our modified code.

The orbit precession showed in Fig. \ref{figyuk} is a real physical effect, it is not related to any numerical instabilities. YGP decreases with distance more steeply than the Newtonian potential. It is well known, from the Bertrand's theorem, that potentials that fall with distance differently from $1/r$ make the orbits to precess; in other words, the orbits are not closed, as we show in Fig. \ref{figyuk}. In the following sections, we will study this issue in detail.

\subsubsection{Timesteps in the two-body simulation}

From Equation \ref{timestep}, the $\Delta t$ values depend strongly on the acceleration of the particles. But the acceleration depends also on some other parameters, for example, the softening length $\epsilon$ and the YGP parameter $\lambda$. When the distance between two particles is smaller than the softening length scale, the gravity law in the simulation is replaced by a softening kernel that avoids divergences in the calculated field (see, e.g., \cite{springel2005}). Due to the numerical nature of our calculations, the investigation of the timestep values is a very important task, due to the fact that their values are used by the KDK algorithm to update positions and velocities. Therefore, it is necessary to probe the influence of the numerical parameters $\lambda$ and $\epsilon$ on the values of the timesteps. In this sense, we have rerun the two-body simulations described above, changing the values of the parameters $\lambda$ and $\epsilon$. Concerning the $N$-body system, we will show in the next sections why \textbf{Gadget-2} is a safe code to deal with these parameters, displaying the $\Delta t$ values for different parameters $\lambda$ and $\epsilon$.

In Table \ref{table1}, we display the maximum $\Delta t$ values from a set of 6 runs, where we have set the same simulated time and used the same two-body system described in Subsection \ref{binary}. We have only changed the values of $\epsilon$ and $\lambda$,  setting $\eta=0.025$ for all experiments, to improve the timestep accuracy.

In these simulations, we run the two-body system for 6 Gyears for  $\lambda=$ 1, 10 and 100 kpc, and
with $\epsilon$ = 0.1 and 1 kpc. Although setting $\epsilon$ values for two-body tests seems to be unnecessary,
we want to probe how these parameters change our results. We have also made a newtonian simulation for a two-body system using the default \textbf{Gadget-2}, to compare our modified code to the original one. In the newtonian case, we obtain  $0.0015 \le  \Delta t \le 0.0031$ for $\epsilon=1$ kpc and $0.00038 \le  \Delta t \le 0.00076$ for $\epsilon=0.1$ kpc.

It is worth mentioning that for all the runs, for different values of $\lambda$, $\epsilon$ and timesteps,
the total energy conserves very well, as we will see below.

\begin{table}
\caption{Some runs with  different $\epsilon$ and $\lambda$ values.  The first column shows the $\lambda$ values, the second one the $\epsilon$ values and the third one, $\Delta t$. The first line display the results of the simulation described in Subsection \ref{binary}.}
\label{table1}
\centering
  \begin{tabular}{@{}lcc@{}}
  \hline\hline
 $\lambda$ & $\epsilon$ & $\Delta t$  \\
\hline
100.0 & 1.0 & $0.0015 \le  \Delta t \le 0.0059$   \\
100.0 & 0.1 &  $0.0003 \le  \Delta t \le 0.0036$  \\
10.0  & 1.0 &  0.0059  \\
10.0  & 0.1 &  0.0059 \\
1.0   & 1.0 &  0.0059 \\
1.0   & 0.1 &  0.0059 \\
\hline
\end{tabular}
\end{table}

\subsubsection{Testing the reliability of our simulations}
\label{testing}

A good integrator must conserve the total energy of the simulated system within an acceptable level. However, due to the nature of the YGP, orbits are not closed and this feature reveals itself as a good indicator to probe how reliable is our two-body simulation. So,  we performed four additional tests to verify the reliability of our simulation described in the Subsection \ref{binary}, namely: energy conservation, angular momentum, orbital period and precession of the orbit.

\begin{figure}
\begin{center}
\includegraphics[width=10cm]{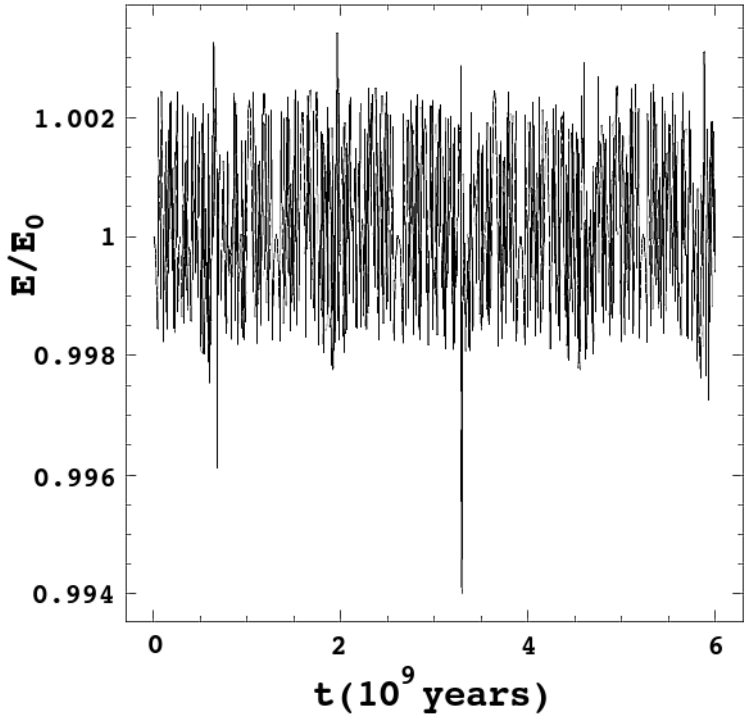}
\end{center}
\caption{The total energy (normalized by the total initial energy $E_0$) evolution as function of time for the Newtonian case.}
\label{newtenerg}
\end{figure}

\begin{figure}
\begin{center}
\includegraphics[width=10cm]{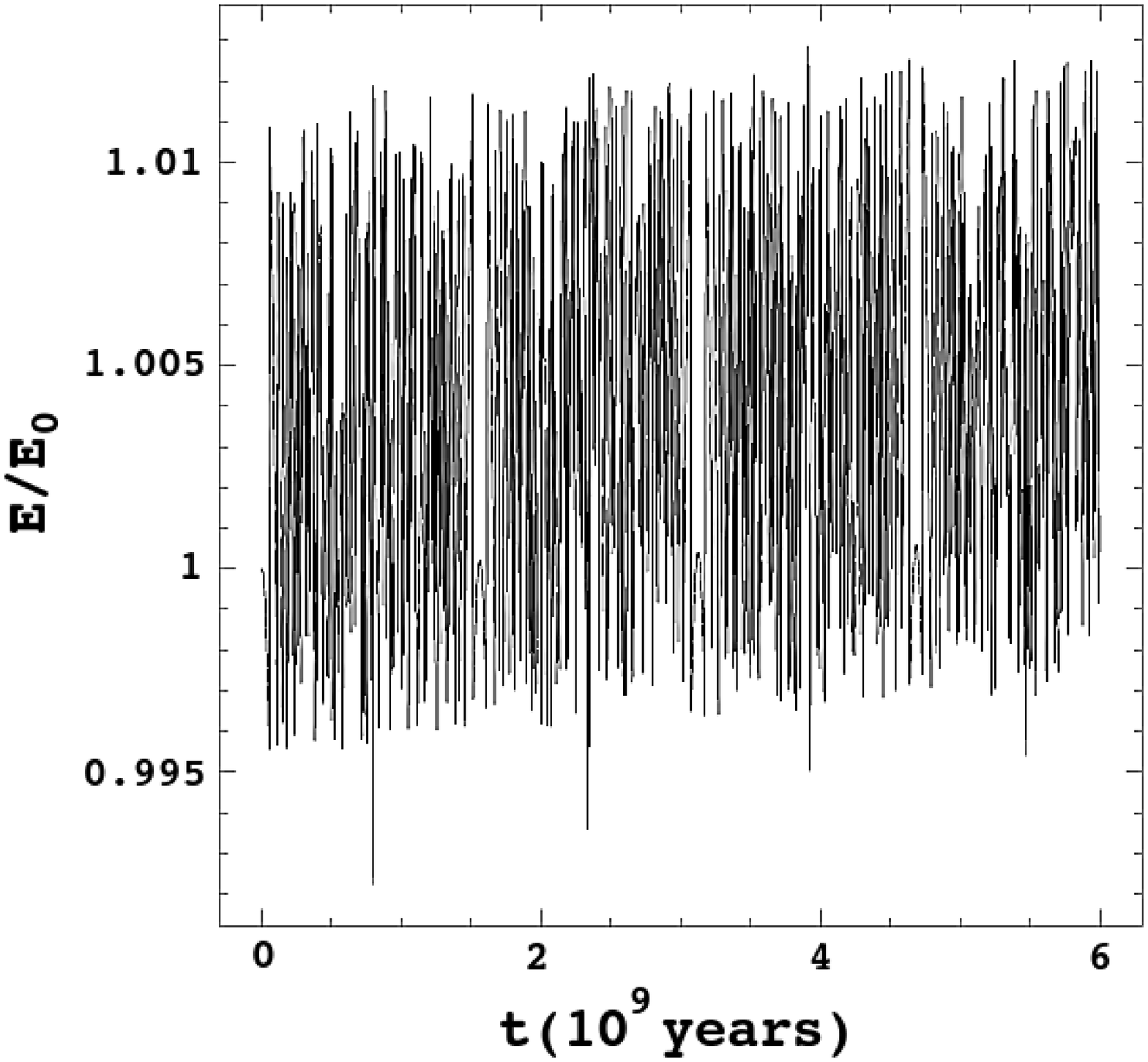}
\end{center}
\caption{The total energy (normalized by the total initial energy $E_0$) evolution as function of time for the Yukawian case with $\lambda=100$ kpc.}
\label{yukenerg}
\end{figure}

In the energy conservation test, we compare our results with the Newtonian case. Thus, we have made another simulation with the original (Newtonian) code. This simulation is performed with the same initial conditions, as we explained in the last section. In Figs. \ref{newtenerg} and \ref{yukenerg}, we show the total energy evolution for both simulations. These pictures show the evolution of the total energy $E$ normalized by the total initial energy $E_0$.  To the cases where $\lambda=100$ kpc, we consider $\epsilon=0.1$ and $\epsilon=1$ and the results are practically the same. Therefore, we display here only the results for $\epsilon=1$ and $\lambda=100$ kpc.

In Fig. \ref{newtenerg}, we conclude that the maximum energy violation $(E-E_0)/E_0$ is about $0.6 \%$ at $t \simeq 3.3$ Gyears. In the YGP case, as can be seen in Fig. \ref{yukenerg}, the energy violation is systematically $(E-E_0)/E_0 \simeq 1.3 \%$. Note that the energy conservation is better in Newtonian case, this has to due with the fact that $1/r$ behaves numerically better than $e^{-r/\lambda}$.
However, we consider that this violation can be neglected for two reasons: firstly, 6 Gyears is a great time lapse for such an energy violation. Secondly, this violation is systematic, at some snapshots the violation is null and this behavior also appears in Newtonian case. It is interesting to note that at 6 Gyears the energy violation is almost null in both situations, showing that our method is consistent.

One could ask what happen with the energy conservation if we follow the simulations, say, till 100 Gyears.
The results of a such simulation show that the energy conservation is systematically small, just like in the
6 Gyears simulation. Such a result shows again the reliability of our code.

\begin{figure}
\begin{center}
\includegraphics[width=10cm]{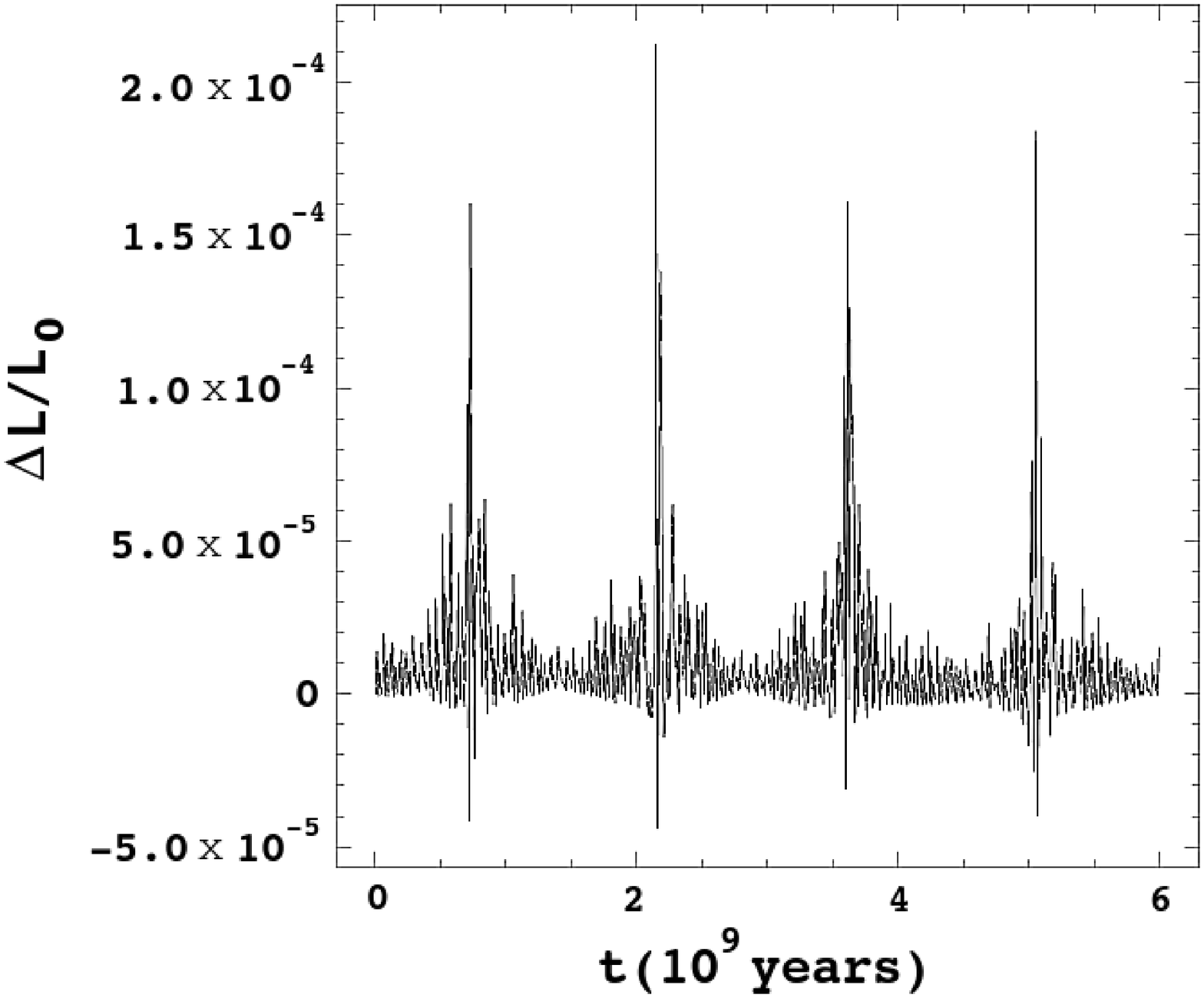}
\end{center}
\caption{Evolution of the angular momentum (normalized by the initial angular momentum) for $\lambda=100$ kpc simulation.}
\label{yukangmom}
\end{figure}

We have also tested the conservation of angular momentum. Fig. \ref{yukangmom} shows  the evolution of the total angular momentum $L$ normalized by the total initial angular momentum $L_0$. We note that the total angular momentum is violated by an amount of 0.2 $\%$ over the initial value.

Besides the energy and the angular momentum conservation, we have also performed three additional tests.
The first one is to verify the orbital half-period of the test particle. The second one is to check the
precession of the particle's orbit. The third test investigates the difference between the potentials
calculated by the tree and the direct sum methods in the modified code. To do this test, we use 30,000
particles to represent a typical snapshot of galaxy simulations.

\begin{figure}
\begin{center}
\includegraphics[width=10cm]{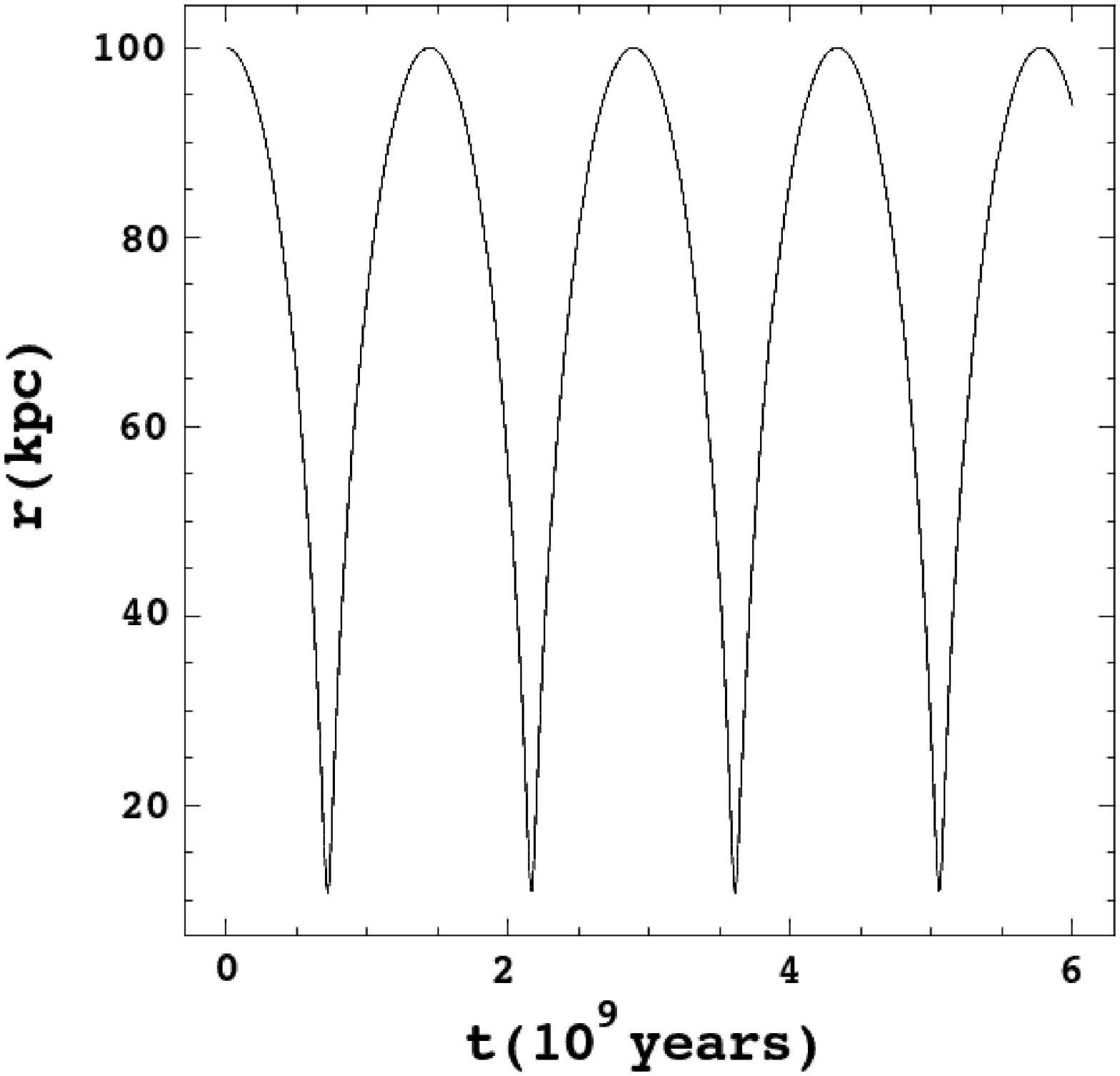}
\end{center}
\caption{Evolution of the particle's distance from the center of mass frame for the $\lambda=100$ kpc simulation.}
\label{distevo}
\end{figure}

\begin{figure}
\begin{center}
\includegraphics[width=10cm]{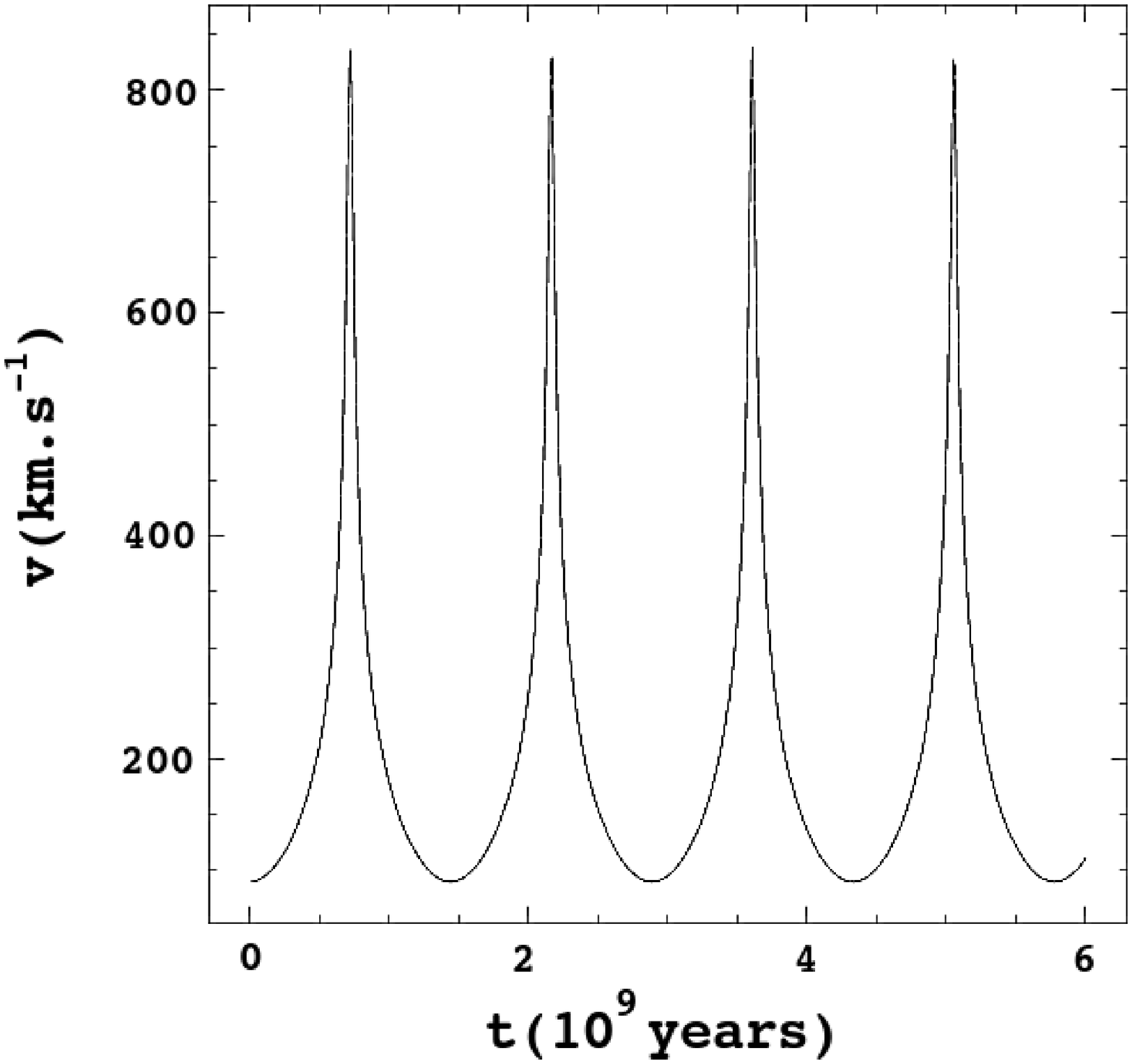}
\end{center}
\caption{Evolution of the particle's velocity from the center of mass frame for the $\lambda=100$ kpc simulation.}
\label{velevo}
\end{figure}

In the test of the period, we evaluate the half-period, which is given by:

\begin{equation}
T_r =  \int_{r_{ap}}^{r_{p}} \frac{dr}{\sqrt{2[E - \Phi (r)] - \frac {L^2}{r^2}}},
\label{orbper}
\end{equation}
where $T_r$ is the orbital half-period, $r_{ap}$ is the apocenter, $r_p$ is the pericenter, $E$ is the initial total energy per unit mass, $\Phi(r)$ is the potential over the test particle, $r$ is the distance from the center of mass to the test particle and $L$ the angular momentum per unit mass. We calculate the above integral numerically and found a value of 0.78 Gyears. In our simulation, the numerical value obtained from the snapshots is also 0.78 Gyears, i.e., the results agree quite well. This result can also be seen in Figs. \ref{distevo} and \ref{velevo}.

The second test consists in the evaluation of the precession angle. Bertrand's theorem states that potentials decreasing with distance differently from $1/r$ make the orbits to precess (see, e.g., Goldstein \cite{goldstein}). We then evaluated the precession using the following equation:

\begin{equation}
\Delta \varphi= 2 L \int_{r_{ap}}^{r_{p}} \frac{dr}{r^2 \sqrt{2[E - \Phi (r)] - \frac {L^2}{r^2}}},
\label{precessaoeq}
\end{equation}
where $\Delta \varphi$ is the shift of the semi major axis. We again evaluate this integral numerically and found that $\Delta \varphi = 28.72^\circ$. This value agrees with the results obtained from a snapshot of our simulation. Taking  the snapshot at t=0.78 Gyears, we found that the particle's position, from the first snapshot, is $\Delta \Phi= 28.72^{\circ}$.  This result again shows that our YGP \textbf{Gadget-2} works quite well and reproduces many physical properties of this kind of potential, due to the good accuracy provided by its integrator's algorithm.

\subsection{Tests with $N$-body systems}

So far we have investigated the two-body problem, whose tests demonstrated that our modified code produces reliable results. Now, we consider the $N$-body problem, for which \textbf{Gadget-2} was designed. In this way, we are initially interested in the recursive octaltree algorithm accuracy to compute the YGP, when compared to the particle-particle method. We are also interested in the influence of the numerical parameters $N$, $\lambda$ and $\epsilon$ over the timestep values, as well as in the energy conservation. In what follows, we will show the main results of a set of $N$-body simulations.

\subsubsection{The recursive tree algorithm in the YGP case}

These tests consist in the comparison of the gravitational potentials of a system of particles calculated by two different methods: the exact one, also called the direct sum method or particle-particle (PP) method, that computes all potentials using $\sim N^2$ operations, and the treecode method, used by \textbf{Gadget-2} code, as we explained above. In this test, we use both codes, the default  \textbf{Gadget-2} and our modified version. To do this, we have chosen a typical snapshot used in galactic dynamics. This snapshot describes a system of particles modeling the initial conditions of one exponential disk galaxy composed by 10,000 particles embedded in a Hernquist dark matter halo composed by 20,000 particles. Both systems are built from typical parameters, like the galaxy models developed by  Springel \& White \cite{springel1999}. Later on we will see that such a resolution is enough to test the reliability  of the modified treecode method to compute the YGPs of all the particles.

\begin{figure}
\begin{center}
\includegraphics[width=10cm]{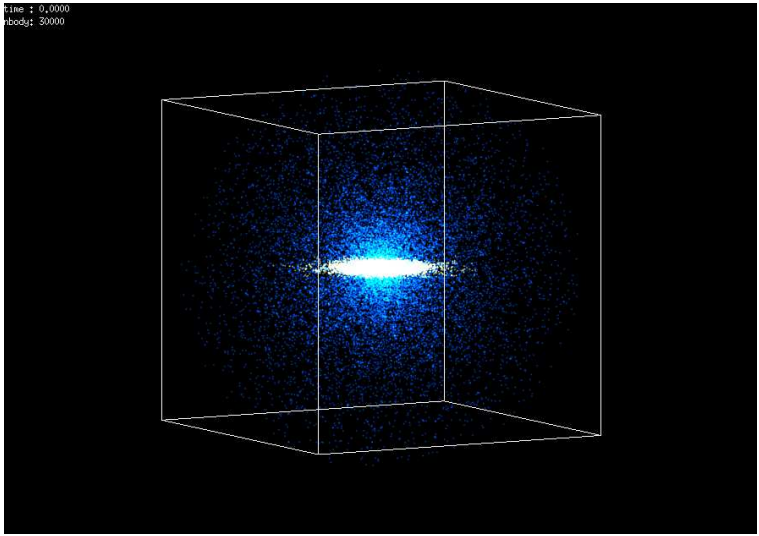}
\end{center}
\caption{Typical snapshot modeling a spiral galaxy and its dark matter halo seen in perspective. Each box side has 80 kpc. Blue particles represent dark matter halo and white/yellow particles indicate the disk.}
\label{snapshot}
\end{figure}

In Fig. \ref{snapshot} we show a visualization of our galaxy model. In another publication to appear elsewhere we show such galaxy models in detail.

Using a Fortran code, we have computed the total potential of each halo and disk particles by the direct sum method and compared these results to the potentials computed by \textbf{Gadget-2} code. We define $ \Delta \phi = |\phi_{Nbody}-\phi_{Tree} |$ as the modulus of the difference resulting from computation of the potential by the two different methods. $\phi_{Nbody}$ indicates the potential calculated by the direct sum method with our Fortran code, and $\phi_{Tree}$, the potential calculated with the \textbf{Gadget-2}, that uses the tree method. To do this, we have used typical parameters of the tree, e.g., the tolerance parameter $\theta=0.8$, and softening lengths of $\epsilon_h=1$ kpc for the dark matter halo and $\epsilon_d=0.4$ for the baryonic disk. These parameters are unchanged in the Newtonian and the YGP cases, i.e., they are fixed parameters. Defining the relative error $\Delta \phi / \phi_{Nbody}$, we can compare both methods. The system of units used here is $10^{10} M_{\odot} $ for unit mass, kpc for distances and $\rm{km.s^{-1}}$  for velocities.

\begin{figure}
\begin{center}
\includegraphics[width=10cm]{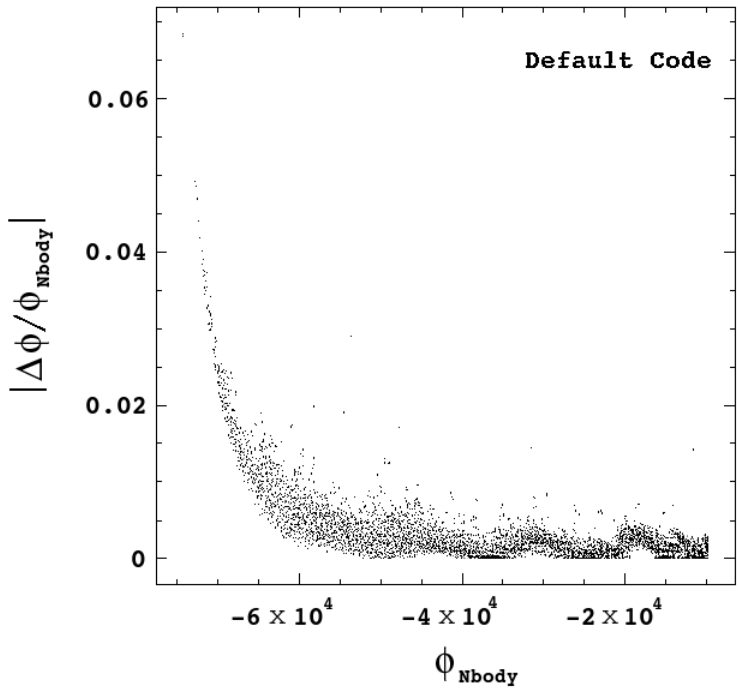}
\end{center}
\caption{Modulus of the relative errors as a function of the gravitational newtonian potential for the particles of the dark matter halo and for the baryonic disk.}
\label{deltaphi_n}
\end{figure}

\begin{figure}
\begin{center}
\includegraphics[width=10cm]{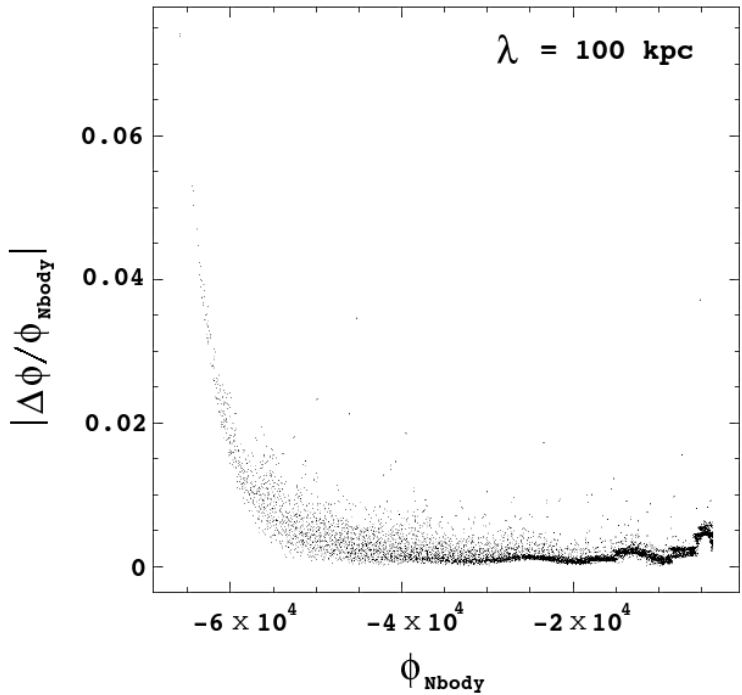}
\end{center}
\caption{The same as Figure \ref{deltaphi_n} for YGP with $\lambda=$ 100kpc.}
\label{deltaphi_100}
\end{figure}

In Fig. \ref{deltaphi_n}, we display $\Delta \phi / \phi_{Nbody}$ against the potentials of the particles in the newtonian case, while in Fig. \ref{deltaphi_100} we show the YGP case, having in mind that both cases are calculated to the same snapshot. To compute the potentials in the YGP case, we have set $\lambda$=100 kpc, the same halo's length displayed in Fig. \ref{snapshot}. Due to the fact that the halo and $\lambda$ have equal length in our snapshot, the $e^{-r/\lambda}$ factor plays a important role in the potential computations at medium and great separations between the particles. Note that if two particles are at distance of 2.5 kpc, the potential is 0.976 Newtonian (see Eq. \ref{yukpot}). But, in the case where the distance is 10 kpc, the Yukawian potential is 0.9 of the Newtonian, and when the distance is 25 kpc, the potential is 0.78 of the Newtonian.

Analyzing Figs. \ref{deltaphi_n} and \ref{deltaphi_100}, we conclude that in the YGP case for $\lambda = 100$ kpc
the tree precision behaves very similarly to the Newtonian case. Errors are the same in both cases and our figures show that the modified code can compute potentials of $N$-body systems with the same precision as the Default Code. Note that we have used the initial snapshots in Figs. \ref{deltaphi_n} and \ref{deltaphi_100}, but similar
results are obtained for any other snapshots of our simulations.

\subsubsection{Additional $N$-body tests: $\lambda$ = 1 kpc and $\lambda=$ 10kpc.}

In addition to the tests described above, we make simulations now considering $\lambda=1$ and 10 kpc, using a snapshot similar to the previous one, but with a better resolution, namely: 30,000 particles for the disk and for the dark halo, totalizing 60,000 particles. In this new test, we have changed the $\epsilon$ values to $\epsilon_h=0.1$ and $\epsilon_d=0.05$. Our computational procedures are very similar to that described in the previous Subsection and here we show that $\phi_{tree}$ and $\phi_{Nbody}$ are practically the same.

\begin{figure}
\begin{center}
\includegraphics[width=10cm]{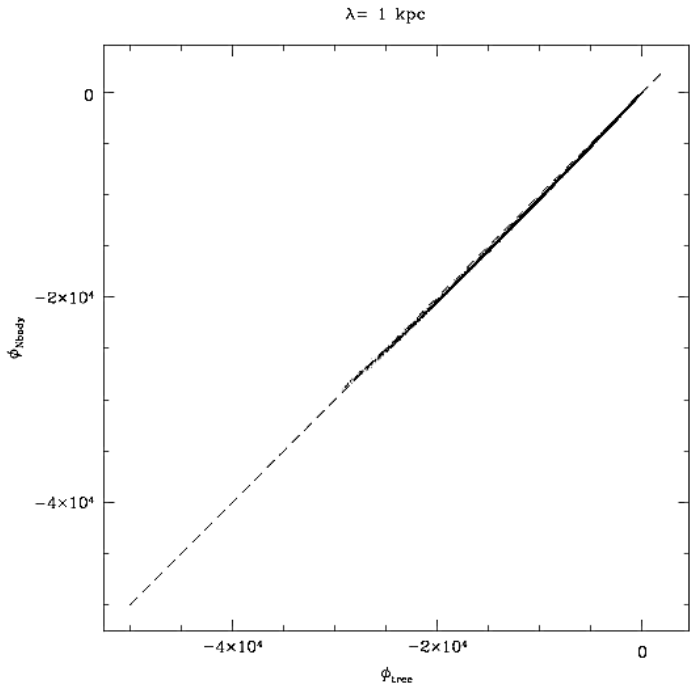}
\end{center}
\caption{The octtree accuracy for $\lambda=1$ kpc case. $\phi_{tree}$ vs. $\phi_{Nbody}$ for all the particles of our simulations; the dashed line represents $\phi_{Nbody}=\phi_{tree}$.}
\label{phis_1}
\end{figure}

\begin{figure}
\begin{center}
\includegraphics[width=10cm]{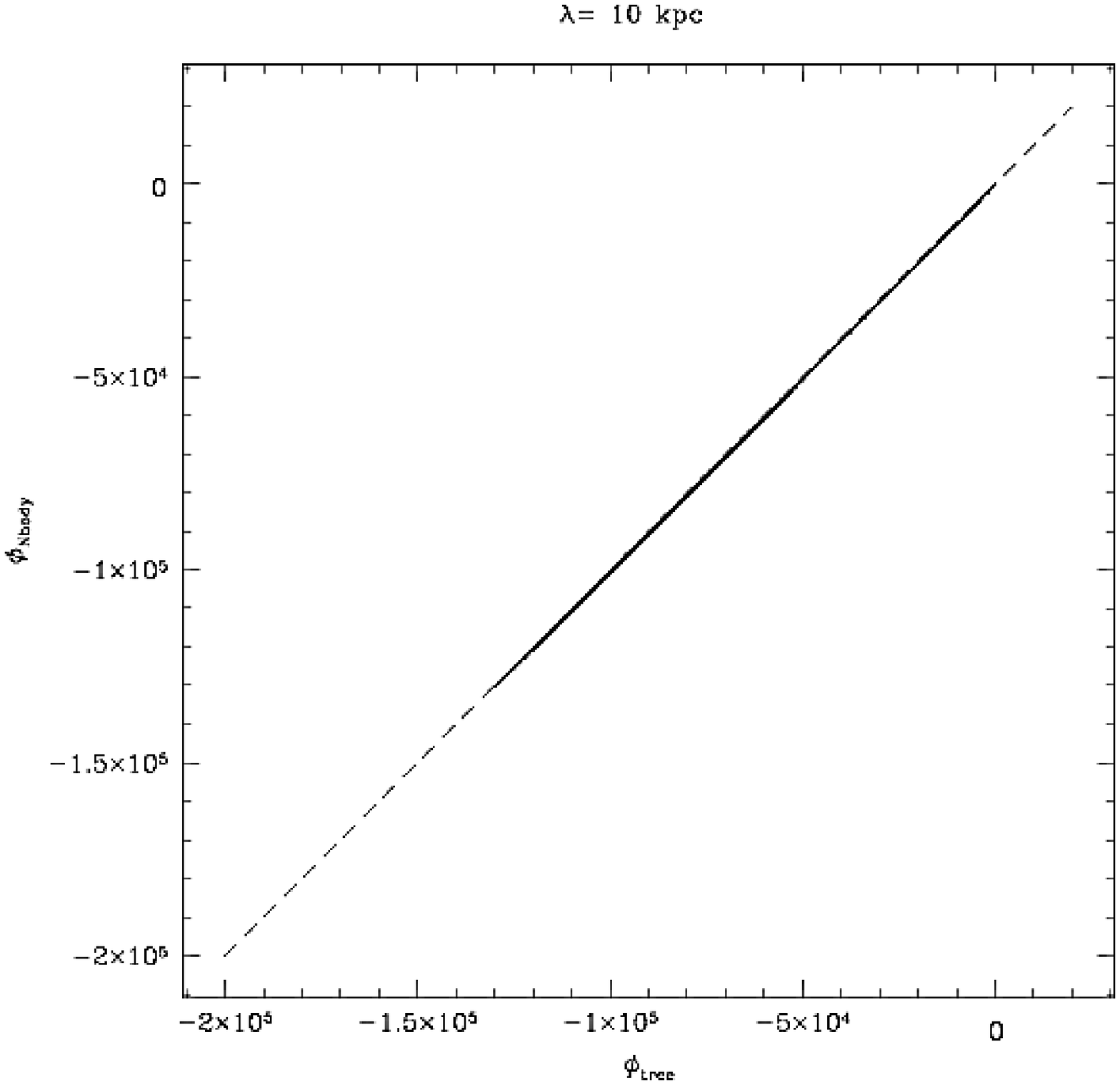}
\end{center}
\caption{The octtree accuracy for $\lambda=10$ kpc case. $\phi_{tree}$ vs. $\phi_{Nbody}$ for all the particles of our simulations; the dashed line represents where $\phi_{Nbody}=\phi_{tree}$.}
\label{phis_10}
\end{figure}

The results are displayed in the Figs. \ref{phis_1} and \ref{phis_10}, where we have plotted $\phi_{tree}$ against $\phi_{Nbody}$ for all the particles of our simulations; the dashed line represents $\phi_{Nbody}=\phi_{tree}$. Note that the points are distributed along the dashed line, with a very small scattering. This result shows that our modified code calculates very well the gravitational potential.

\subsubsection{Timestep tests}

Numerical simulations are subject yet to considerable controversy, mostly concerning their numerical collisionless nature and their resolution. It is well known that the parameters $N$, $\epsilon$ and $\Delta t$ can change the simulation's results, and, in the galactic dynamic scenarios, the final morphology can depend on this set of parameters.

An important detailed study in this issue is provided by Gabbasov et. al.\cite{gabbasov}, where they probe the influence of the numerical parameters $N$, $\epsilon$ and $\Delta t$ on the outcomes of a set of simulations with disk galaxies and issues related to the formation of bars in these late-type systems. They prove, for example, that different sets of these parameters can change significantly the results of simulations that have initially the same snapshot.

In this paper, we make some tests similar to those by Gabbassov's team to probe the influence of the numerical parameters $N$, the total number of particles, $\epsilon$, and $\lambda$ on the timesteps values, using a methodology similar to that used in \cite{gabbasov}.  It is important to bear in mind that our tests are somewhat less complex and we performed a set of numerical simulations up to 0.1 Gyears of simulated time only to verify the influence of these parameters.

This simulated time is short, if compared to the many runs performed by \cite{gabbasov}. However, due to the \textbf{Gadget-2} method in computing the timesteps, each of our runs take $\sim$ 500-5000 steps to be completed, being a very satisfactory number of steps to our purposes.

Besides, in some tests performed with 30,000 and 300,000 particles over 1 Gyear, the $\Delta E / E_0$ values did not change significantly during all the simulated time, although 50,000 steps were necessary to this end.

In this sense, 0.1 Gyears are sufficient for our studies. In this way, we have used three snapshots with 30,000, 300,000 and 1,200,000 particles, conserving the ratio 1:2 for the baryonic disk and dark halo, respectively. We run these snapshots for different values of  $\epsilon$ and $\lambda$. These snapshots represent an exponential disk embedded in a Hernquist Sphere, as we explained above.

\begin{table}
 \caption{Some runs for different $N$, $\epsilon$ and $\lambda$ values (in kpc). Note that the respective $\Delta t$ values are calculated by the \textbf{Gadget-2} code. The energy conservation $\Delta E / E_0$ is also shown for each run.}
\label{table2}
\centering
  \begin{tabular}{@{}lccccc@{}}
  \hline\hline
 $N$ & $\lambda$ & $\epsilon_h$ & $\epsilon_d$ & $\Delta t$  & $\Delta E / E_0 \le $ \\
\hline\hline
30\,000      & 1.0 & 1.0  &  0.4  & 0.0016 & $ 1 \% $ \\

30\,000      & 1.0 & 0.1  & 0.04  & $0.00020 \le  \Delta t  \le  0.00039$ & $ 0.3 \%$ \\

300\,000     & 1.0 & 0.1  & 0.04  & $0.00020 \le  \Delta t  \le  0.00078$ & $   0.015 \%$  \\

300\,000     & 1.0 & 0.01 & 0.004 & $4.88281 \times 10^{-5}$ & $ 0.1 \%$  \\

1\,200\,000 & 1.0 & 0.1  & 0.04  & $0.00020 \le  \Delta t  \le  0.00039$  & $ 0.01 \%$  \\

1\,200\,000 & 1.0 & 0.01 & 0.004 & $9.76562 \times 10^{-5}$  & $ 0.02 \%$  \\

\hline

30\,000      & 10.0 & 1.0  & 0.4   & 0.00078 & $ 0.4 \%$ \\

30\,000      & 10.0 & 0.1  & 0.04  & 0.00020 & $ 0.15 \%$ \\

300\,000     & 10.0 & 0.1  & 0.04  & 0.00020 & $ 0.02 \%$  \\

300\,000     & 10.0 & 0.01 & 0.004 & 2.44141 $\times 10^{-5}$ & $ 0.15 \%$  \\

1\,200\,000 & 10.0 & 0.1  & 0.04  & 0.00020 & $ 0.015 \%$  \\

1\,200\,000 & 10.0 & 0.01 & 0.004 & 4.88281 $\times 10^{-5}$ & $ 0.04 \%$  \\

\hline

30\,000      & 100.0 & 1.0  & 0.4   & 0.00078 & $ 0.08 \%$ \\

30\,000      & 100.0 & 0.1  & 0.04  & 0.00020 & $ 0.08 \%$  \\

300\,000     & 100.0 & 0.1  & 0.04  & 0.00020 & $ 0.06 \%$   \\

300\,000     & 100.0 & 0.01 & 0.004 &  2.44141 $\times 10^{-5}$  & $ 0.2 \%$  \\

1\,200\,000 & 100.0 & 0.1  & 0.04  &  0.00020 & $ 0.04 \%$  \\

1\,200\,000 & 100.0 & 0.01 & 0.004 & 4.88281 $\times 10^{-5} $ & $ 0.05 \%$  \\
\hline
\end{tabular}
\end{table}

To make the runs displayed in the Table \ref{table2}, we estimate the $\epsilon$ values taking the mean inter-particle separation at the center of the halo and disk. Our values are near to those used by \cite{gabbasov} and are typical, if compared to those cited in the literature. Note that in the cases where $\epsilon < 0.01$, the simulations gets ``more collisional" and the energy conservation gets a little bit worse, in agreement to the results obtained by \cite{gabbasov}.  If the resolution is increased, in other words, if the $\epsilon$ values are smaller, then timestep values are smaller as well. As a consequence, the \textbf{Gadget-2} tree algorithm computes greater accelerations and potentials at these smaller scales than in the small resolution cases (greater $\epsilon$ values).

However, in these cases where $N$ is very large, the system tends to be collisionless, because in the \textit{close encounters}, the softening kernel prevent divergences of accelerations and potentials, once $\epsilon$ is chosen appropriately. Also, the collective potential dominates over individual interactions, making results better to $N > 10^5 $ particles.

So, the greater the resolution is (i.e., the greater N) the better is the energy conservation as long as the $\epsilon$ values are correctly chosen.
But, note that, even with a modest value of N, the energy conservation can be quite good. Also, to probe
a given alternative theory it is not always necessary a high resolution simulation, since the global properties
of a galaxy, such as, for example, the density profile, can be very sensitive to the alternative potential. In other papers to appear elsewhere we consider this issue for early and late type galaxies.

From our results, we consider our modified code able to test the YGP.  Therefore, we are able to study alternative theories using galaxies or globular clusters, with the same quality one obtains with the use of Newtonian codes.

\section{Conclusions}

The present paper considers the use of the well known \textbf{Gadget-2}, with the appropriate modifications, to study alternative theories of gravitation, with particular attention to the YGP. In this first paper we discuss how to modify the \textbf{Gadget-2} and evaluate how reliable this modified code is, performing a series of tests.

As we have shown in the previous sections, our tests are successful: numerical errors from our changes on \textbf{Gadget-2} are neglected and we can use this modified code to probe alternative theories of gravitation. For example, we can use arguments based on galactic dynamics to exclude or not the Yukawa potential hypothesis. This is very important, due to the fact that alternative theories appear frequently in the literature, but so far there are no many investigations in  the galactic scales using $N$-body codes and galactic dynamics formalism. We will show, in future papers, how to test these alternative theories using early and late-type galaxies.

\begin{acknowledgements}
The authors would like to thank the Brazilian agencies CAPES, CNPq and FAPESP for support. The authors would like also to thank M\'arcio Eduardo da Silva Alves for interesting suggestions, discussions and for the critical reading of the paper.
\end{acknowledgements}

\end{document}